\documentclass[useAMS,usenatbib]{mn2e}

\usepackage{epsfig}
\usepackage{amsmath}
\usepackage{amssymb}
\usepackage{natbib}
\usepackage{threeparttable}

\newcommand{\exo}{EXO~0748--676 }
\newcommand{\chan}{\textit{Chandra }}

\newcommand{\swift}{\textit{Swift }}
\newcommand{\xmm}{\textit{XMM-Newton }}

\def \aj {AJ}
\def \mnras {MNRAS}
\def \apj {ApJ}
\def \apjs {ApJS}
\def \apjl {ApJL}
\def \aap {A\&A}

\def \atel {ATel}

\def \pasj {PASJ}
\def \aaps {AAPS}

\def \pre {PhRvE}

\title[Chandra and Swift observations of \exo back to quiescence]{Chandra and Swift observations of the quasi-persistent neutron star transient \exo back to quiescence}
\author[N. Degenaar et al.]
{N. Degenaar$^{1}$,
R. Wijnands$^{1}$, 
M.T. Wolff$^{2}$, 
P.S. Ray$^{2}$, 
K.S. Wood$^{2}$, 
J. Homan$^{3}$, 
\newauthor W.H.G. Lewin$^{3}$, 
P.G. Jonker$^{4,5}$, 
E.M. Cackett$^{6,7}$, 
J.M. Miller$^{6}$, 
E.F. Brown$^{8}$\\ 
$^{1}$Astronomical Institute "Anton Pannekoek", 
University of Amsterdam, 
Kruislaan 403, 1098 SJ, Amsterdam, the Netherlands\\
$^{2}$Space Science Division, 
Naval Research Laboratory, 
Washington, DC 20375, USA\\
$^{3}$MIT Kavli Institute for Astrophysics and Space Research, 
70 Vassar Street, Cambridge, MA 02139, USA\\
$^{4}$SRON, Netherlands Institute for Space Research, 
Sorbonnelaan 2, 3584 CA, Utrecht, the Netherlands\\
$^{5}$Harvard-Smithsonian  Center for Astrophysics, 
60 Garden Street, 
Cambridge, MA~02138, U.S.A.\\
$^{6}$University of Michigan, 
Department of Astronomy, 
500 Church Street, Dennison 814, Ann Arbor, MI 48105, USA\\
$^{7}$Chandra fellow\\
$^{8}$Department of Physics and Astronomy, 
Michigan State University, 
East Lansing, MI 48824, USA
}

\begin{document}

\date{Received 27 November 2008 / Accepted 10 March 2009}

\pagerange{\pageref{firstpage}--\pageref{lastpage}} \pubyear{0000}

\maketitle

\label{firstpage}

\begin{abstract} 
The quasi-persistent neutron star X-ray transient and eclipsing binary \exo recently started the transition to quiescence following an accretion outburst that lasted more than 24 years. We report on two \chan and twelve \swift observations performed within five  months after the end of the outburst. 
The \chan spectrum is composed of a soft, thermal component that fits to a neutron star atmosphere model with $kT^{\infty} \sim 0.12$~keV, joined by a hard powerlaw tail that contributes $\sim 20\%$ of the total 0.5-10 keV unabsorbed flux. The combined \textit{Chandra/Swift} data set reveals a relatively hot and luminous quiescent system with a
temperature of $kT^{\infty} \sim0.11-0.13$~keV and a bolometric thermal luminosity of $\sim 8.1 \times 10^{33}-1.6 \times 10^{34}~\mathrm{(d/7.4~kpc)^2~erg~s}^{-1}$. We discuss our results in the context of cooling neutron star models.  
\end{abstract}


\section{Introduction}
Neutron star X-ray transients spend the vast majority of their time in quiescence, in
which they are dim with typical luminosities of
$10^{32-34}$~erg~s$^{-1}$, but occasionally show an immense X-ray
brightening in which their luminosity can rise to levels of $10^{36-38}$~erg~s$^{-1}$ \citep[e.g.,][]{chen97}. Their quiescent X-ray spectra are observed to
consist of one or two components; a soft, thermal component ($kT \sim
0.1-0.2$~keV), and/or a hard power-law tail \citep[dominating above 2 keV, photon index
1-2; e.g.,][]{asai1996}. 

Several explanations have been put forward to describe the quiescent emission of neutron star transients, such as low-level accretion \citep[e.g.,][]{zampieri1995,menou99} or emission mechanisms connected to the magnetic field of the neutron star \citep[see e.g.,][]{campana2003}. However, the soft spectral component is most often interpreted as thermal emission emerging from the neutron star surface \citep{brown1998}. During accretion outbursts, a series of nuclear reactions deposit heat in the neutron star crust \citep[e.g.,][]{haensel1990b,haensel2008,gupta07}, which spreads over the neutron star. The gained heat is radiated as thermal emission from the surface once the system returns to quiescence. In this interpretation the quiescent thermal emission depends on the time-averaged accretion rate of the system \citep[e.g.,][]{brown1998}, as well as on the neutrino emission mechanism that operates in the core, which regulates the cooling \citep[e.g.,][]{yakovlev03}. 

There exists a small group of quasi-persistent X-ray transients, which undergo prolonged accretion outbursts with a duration of years to decades rather than the usual weeks to months \citep[e.g.,][]{wijnands2004_review}. In these systems, the neutron star crust is substantially heated and becomes thermally decoupled from the core. Once the outburst ends, the crust will cool down primarily through heat conduction towards the core, until eventually thermal equilibrium is re-established \citep[e.g.,][]{rutledge2002}. This thermal relaxation depends strongly on the properties of the crust, such as the thermal conductivity.

In recent years, the quasi-persistent X-ray binaries KS~1731--260 and
MXB~1659--29 have been monitored during the transition towards quiescence with \chan and \xmm
following accretion outbursts with a duration of $\sim 12.5$ and $\sim 2.5$ yrs respectively
\citep[][]{wijnands2001,wijnands2002,wijnands2003,wijnands2004,cackett2006,cackett2008}.
For both systems, these observations revealed a lightcurve
that decayed exponentially from a bolometric luminosity of $\sim (3-5)
\times 10^{33}~\mathrm{erg~s}^{-1}$ a few months after the outburst,
leveling off to $\sim (2-5) \times 10^{32}~\mathrm{erg~s}^{-1}$
several years later \citep[][]{cackett2006,cackett2008}. Whereas the initial stages of this decaying curve are set by the properties of the crust, the quiescent base
level reflects the thermal state of the core \citep[e.g.,][]{brown08}. 
Confronting the observed cooling curves with thermal evolution models suggests that the neutron stars in both KS~1731--260 \citep[][]{shternin07} and MXB~1659--29 \citep{brown08} have a highly conductive crust. This idea is supported by theoretical plasma simulations of \citet{horowitz2007}, who demonstrated that the accreted matter will arrange itself in a lattice structure with a high thermal conductivity. 

\subsection{\exo}\label{subsec:exo}
Recently, the quasi-persistent neutron star X-ray transient EXO~0748--676 also started the transition to quiescence. This low-mass X-ray binary was initially discovered with \textit{EXOSAT} in February 1985 \citep{parmar1986}, although it appears as an \textit{EXOSAT} slew survey source several times before this date \citep[][the earliest detection dates back to 1984 July 15]{reynolds1999}. Prior to its discovery, \exo was serendipitously observed with \textit{EINSTEIN} in May 1980, from which \citet[][]{garcia1999} deduced a $kT\sim$0.2 keV blackbody source spectrum with a 0.5-10 keV luminosity of $\sim5 \times 10^{33}~(\mathrm{d/7.4~kpc})^2~\mathrm{erg~s}^{-1}$. The system displays X-ray dips and exhibits eclipses with a duration of $\sim 8.3$~min every 3.82~hr \citep[][]{parmar1986}. 

Since its discovery, \exo has been consistently detected with luminosities exceeding $10^{36}~\mathrm{erg~s}^{-1}$ by various satellites. In particular, regular monitoring with \textit{RXTE} has shown that the source maintained a relatively steady 2-20 keV flux of $\sim 2 \times 10^{-10}~\mathrm{erg~cm}^{-2}~\mathrm{s}^{-1}$ since 1996 \citep[$L_{\mathrm{X}} \sim 1 \times 10^{36}~(\mathrm{d/7.4~kpc})^2~\mathrm{erg~s}^{-1}$;][]{wolff2008}. However, observations with the Proportional Counter Array (PCA) obtained on 2008 August 12 signaled a decrease in 2-20 keV source flux down to $\sim 7 \times 10^{-11}~\mathrm{erg~cm}^{-2}~\mathrm{s}^{-1}$ \citep[$L_{\mathrm{X}} \sim 5 \times 10^{35}~(\mathrm{d/7.4~kpc})^2~\mathrm{erg~s}^{-1}$;][]{wolff2008}. This decline was confirmed when \textit{Swift} observations with the X-ray Telescope (XRT) performed on 2008 September 28 found the source at a 0.5-10 keV flux of $\sim 2 \times 10^{-12}~\mathrm{erg~cm}^{-2}~\mathrm{s}^{-1}$ (see Table~\ref{tab:swift}, $L_{\mathrm{X}} \sim 1 \times 10^{34}~(\mathrm{d/7.4~kpc})^2~\mathrm{erg~s}^{-1}$). The next set of \textit{RXTE}/PCA observations, carried out on 2008 October 5, failed to detect EXO~0748--676 \citep[][]{wolff2008b}, consistent with the flux level observed with \textit{Swift}/XRT.
Optical and near-IR observations performed in October 2008 of the optical counterpart of EXO~0748--676, UY~Vol, detected a decrease in its optical brightness compared with the active X-ray state \citep[][]{hynes2008,torres2008}.

The observed large decline in X-ray and optical luminosity suggest that \exo is returning to quiescence after having actively accreted for 24 yrs. With its long outburst duration, \exo is a good candidate to look for thermal relaxation of the accretion heated neutron star crust now that the system is returning to quiescence. In this Letter we report on \chan and \swift observations of \exo performed within the first five months after the accretion outburst ceased.

\section{Observations, analysis and results}

\subsection{Chandra data}

As part of our \chan Target of Opportunity (TOO) proposal \exo was observed with \textit{Chandra}/ACIS-S on 2008 October 12-13 22:09-02:51 UTC (obsID 9070) and on 2008 October 15 12:46-17:13 UTC (obsID 10783), for on-source times of $13.8~\mathrm{and}~13.3~\mathrm{ks}$ respectively. We used the \texttt{CIAO} tools (v. 4.0) and standard \chan analysis threads to reduce the data. No background flares were found, so all data were used for further analysis. The ACIS-S3 CCD was operated in a 1/8 sub-array to circumvent any possible pile-up problems. For the resulting frame-time of 0.4 s and the observed fluxes (see Table~\ref{tab:spec}), the pile-up fraction was $< 4 \%$.

Source spectra and lightcurves were extracted from a circular region with a radius of 3\arcsec\ centered on the position of EXO~0748--676. Background events were obtained from an annular region with an inner (outer) radius of 10\arcsec\ (25\arcsec).  The lightcurves of both \chan observations display one eclipse at times consistent with the ephemeris of \citet[][]{wolff2008c}. During the eclipses, only one photon was detected from the source region and a similar amount was found for the normalized background. This indicates that X-rays from the neutron star are not detected during the eclipses. 
To calculate the correct non-eclipse time-averaged fluxes, we reduced the exposure times of the fits files by 500 s \citep[the approximate duration of the eclipses;][]{parmar1986,wolff2008c}. 
Using the FTOOL \texttt{grppha} we re-binned the spectra to contain a minimum of 20 photons per bin.

\begin{figure}
 \begin{center}
          \includegraphics[width=8.2cm]{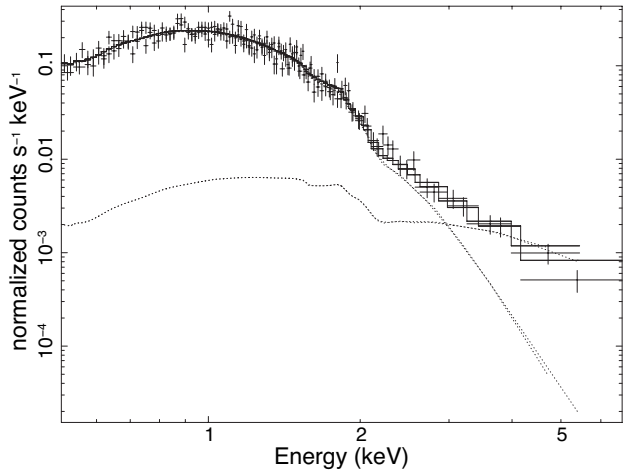}
    \end{center}
\caption[]{{Spectra of both \chan observations, along with the model fit (solid line) for $D=7.4$~kpc, $\Gamma=1$ and $M=1.4~\mathrm{M_{\odot}}$. The dotted lines indicate the \texttt{NSATMOS} and powerlaw components.}}
 \label{fig:spec}
\end{figure}

\begin{table*}
\begin{center}
\caption[l]{Results from fitting the \textit{Chandra}/ACIS-S spectral data.}
\begin{threeparttable}
\begin{tabular}{l c c c c c c c c c}
\hline \hline
& \multicolumn{4}{c}{$\Gamma=1$} & & \multicolumn{4}{c}{$\Gamma=2$}\\
\cline{2-5}  \cline{7-10}\\
Parameter & 5.0 kpc & 5.0 kpc  & $7.4$ kpc & $8.3$ kpc & &5.0 kpc & 5.0 kpc  & $7.4$ kpc & $8.3$ kpc\\
\hline
$N_{\mathrm{H}}~(10^{22}~\mathrm{cm}^{-2})$ & $0.12^{+0.03}_{-0.02}$ & $0.12 \pm 0.02$ & $0.12^{+0.03}_{-0.02}$ & $0.12^{+0.03}_{-0.02}$ & &$0.14 \pm 0.03$ & $0.14^{+0.03}_{-0.02}$ & $0.14^{+0.03}_{-0.02}$ & $0.14 \pm 0.03$ \\
$kT^{\infty}_{\mathrm{eff}}$ (eV) & $112^{+16}_{-12}$ & $113^{+13}_{-10}$ & $118^{+5}_{-3}$ & $119^{+3}_{-4}$ & &$106^{+7}_{-10}$ & $108^{+15}_{-9}$ & $112^{+7}_{-9}$ & $114^{+7}_{-9}$  \\
$M_{\mathrm{NS}}~(\mathrm{M_{\odot}})$ & (1.4) & $1.8^{+0.3}_{-0.5}$ & (1.4) & (1.4) & &(1.4) & $2.02^{+0.2}_{-0.4}$ & (1.4) & (1.4)  \\
$R_{\mathrm{NS}}~(\mathrm{km})$ & $11.9^{+3.0}_{-2.6}$ & (10) & $17.1^{+3.6}_{-2.8}$ & $18.8^{+3.8}_{-3.0}$ & &$14.0^{+3.7}_{-3.0}$ & (10) & $19.4^{+4.7}_{-3.5}$ & $21.3^{+5.1}_{-3.8}$ \\
\hline
$F_{\mathrm{X}}^{\mathrm{}}$ (0.5-10 keV)
& $1.3\pm 0.1$ & $1.3\pm 0.1$ & $1.3^{+0.1}_{-0.2}$ & $1.3\pm 0.1$ & &$1.4\pm 0.1$ & $1.3^{+0.2}_{-0.1}$ & $1.3\pm 0.1$ & $1.3\pm 0.1$ \\
$F_{\mathrm{X}}^{\mathrm{th}}$  (0.01-100 keV)
& $1.5\pm 0.2$ & $1.5^{+0.2}_{-0.1}$ & $1.5\pm 0.1$ & $1.5^{+0.2}_{-0.1}$ &  & $1.5\pm 0.2$ & $1.5^{+0.2}_{-0.1}$ & $1.5\pm 0.2$ & $1.5\pm 0.2$ \\
$L_{\mathrm{X}}$  (0.01-100 keV)
& $4.5\pm 0.6$ & $4.5^{+0.6}_{-0.3}$ & $9.8^{+1.3}_{-0.7}$ & $10.2^{+1.6}_{-0.8}$ &  & $4.5\pm 0.6$ & $4.5^{+0.6}_{-0.3}$ & $9.8\pm 1.3$ &$10.2\pm 1.6$  \\
\hline
\end{tabular}
\label{tab:spec}
\begin{tablenotes}
\item[]Note. -- The quoted errors represent 90\% confidence levels and the reduced $\chi^2$ for all fits is 1.1 (for 173 dof). $F_{\mathrm{X}}$ represents the total unabsorbed 0.5-10 keV flux, while $F_{\mathrm{X}}^{\mathrm{th}}$ gives the bolometric \texttt{NSATMOS} flux (both in units of $10^{-12}~\mathrm{erg~cm}^{-2}~\mathrm{s}^{-1}$). $L_{\mathrm{X}}$ is the bolometric luminosity (for the model distance) of the \texttt{NSATMOS} component in units of $10^{33}~\mathrm{erg~s}^{-1}$. 
\end{tablenotes}
\end{threeparttable}
\end{center}
\end{table*}

The resulting spectra were fitted using \texttt{XSPEC} \citep[v. 12.0;][]{xspec}. The \chan observations were performed  $<3$ days apart, and we did not find any significant spectral changes between the two when fitting the data sets separately. Therefore, we tied all spectral parameters between the two observations. 
We fitted the data with a neutron star atmosphere model \texttt{NSATMOS} \citep{heinke2006}. The normalisation of this model was always fixed to one, which corresponds to the entire neutron star surface emitting. Using only the \texttt{NSATMOS} model, the data above $\sim 2-3$~keV cannot be fit properly.
If we add a powerlaw component, this improves a fit with the neutron star mass and radius fixed at $M_{\mathrm{NS}}=1.4~\mathrm{M_{\odot}}$ and $R_{\mathrm{NS}}=10$~km from $\chi^2=217.73/174~\mathrm{dof}$ to $\chi^2=181.02/172~\mathrm{dof}$ (an F-test suggests a $1.3 \times 10^{-7}$ probability of achieving this level of improvement by chance). The \texttt{NSATMOS} model calculates the effective temperature in the neutron star frame. We converted this to the effective temperature as seen by an observer at infinity  according to $kT^{\infty}_{\mathrm{eff}}= kT_{\mathrm{eff}} \times g_r$, where $g_r = \sqrt{1-(2GM_{\mathrm{NS}})/(R_{\mathrm{NS}}c^2)}$ is the gravitational redshift parameter, $G$ is the gravitational constant and $c$ the speed of light.

When the neutron star mass and radius are fixed to canonical values of $M_{\mathrm{NS}}=1.4~\mathrm{M_{\odot}}$ and $R_{\mathrm{NS}}=10$~km, the best-fit yields $D=3.4^{+1.4}_{-0.7}$~kpc, $\Gamma =2.3\pm0.9$, $N_{\mathrm{H}}=(0.16\pm0.1)\times 10^{22}~\mathrm{cm}^{-2}$ and $kT^{\infty}_{\mathrm{eff}}=100^{+14}_{-23}$~eV. The fitted distance is lower than the best estimate of 7.4 kpc (with an allowed range of 5-8.3 kpc), which was inferred from analysis of type-I X-ray bursts (\citealt{galloway2006}, but see \citealt{galloway2008} for possible additional uncertainties). If we keep the mass and radius at canonical values and in addition fix the distance to either D=7.4 kpc or D=8.3 kpc, we obtain  $\Gamma<0$. However, for $D=5~$kpc we obtain $\Gamma=0.7^{+1.6}_{-0.7}$, $N_{\mathrm{H}}=(0.10\pm0.01)\times 10^{22}~\mathrm{cm}^{-2}$ and $kT^{\infty}_{\mathrm{eff}}=118^{+1}_{-3}$~eV. 

Finally, we explored fits with the distance fixed at 5, 7.4 or 8.3 kpc, but with either the mass or the radius left to vary freely (and the other kept at its canonical value). Since the powerlaw slope is not well constrained, this parameter was fixed to $\Gamma=1$ or $\Gamma=2$. 
The free parameters for each fit are then the hydrogen column density ($N_{\mathrm{H}}$), the effective temperature ($kT_{\mathrm{eff}}$), the normalization of the powerlaw component and either the mass ($M_{\mathrm{NS}}$) or radius ($R_{\mathrm{NS}}$). 

We deduced unabsorbed fluxes in the 0.5-10 keV energy band and calculated the bolometric flux of the thermal component by extrapolating the \texttt{NSATMOS} model (using a zero normalization for the powerlaw) for the energy range 0.01-100 keV. The powerlaw contribution to the total 0.5-10 keV unabsorbed flux is $\sim 16-17\%$ for the fits with $\Gamma=1$ and $\sim 19-20\%$ if $\Gamma=2$. The results are summarized in Table~\ref{tab:spec}. 
For $D=7.4$~kpc and $D=8.3$~kpc the fits with the radius fixed at $R_{\mathrm{NS}}=10$~km resulted in neutron star masses of $M_{\mathrm{NS}}>2.5~\mathrm{M_{\odot}}$, i.e., exceeding the causality limit of $2.23~\mathrm{M_{\odot}}$ for a neutron star radius of 10~km; these fits are not listed in Table~\ref{tab:spec}. The spectra of both \chan observations are plotted in Figure~\ref{fig:spec}.

\begin{table*}
\caption{Results from fitting the \textit{Swift}/XRT spectral data.}
\begin{threeparttable}
\begin{tabular}{l l c c c c c c}
\hline \hline
ObsID & Date & Exposure time (ks) & $kT^{\infty}_{\mathrm{eff}}$ (eV) & $F_{\mathrm{X}}$ & $F_{\mathrm{X}}^{\mathrm{th}}$ & $L_{\mathrm{X}}$ ($10^{33}~\mathrm{erg~s}^{-1}$) & Red. $\chi^2$\\
\hline
51300025 & 2008-09-28 & 0.93 & $133^{+8}_{-11}$ & $2.3\pm 0.6$ & $2.4\pm 0.7$ & $16 ^{+5}_{-4}$ & 1.7 (2 dof)\\
31272001* & 2008-10-07 & 1.49 & $128^{+9}_{-12}$  & $1.9\pm 0.6$ & $2.0 ^{+0.7}_{-0.6}$ & $12\pm4$ & 0.1 (2 dof)\\
31272003/4* & 2008-10-29/30 & 5.01 & $119^{+4}_{-5}$  & $1.4\pm 0.2$ & $1.5\pm 0.2$ & $9.4\pm2$ & 1.4 (10 dof)\\ 
31272005* & 2008-11-02 & 4.78 & $115^{+4}_{-5}$  &  $1.1\pm 0.2$ & $1.3\pm 0.2$ & $8.3\pm 1$ & 1.4 (10 dof)\\
31272007 & 2008-11-28 & 3.04 & $121^{+5}_{-7}$  &  $1.5\pm 0.3$ & $1.6\pm 0.3$ & $11\pm 2$ & 1.0 (5 dof)\\
31272008* & 2008-12-05 & 3.40 & $122^{+5}_{-6}$  &  $1.6\pm 0.3$ & $1.7 \pm 0.3$ & $11\pm 2$ & 0.8 (6 dof)\\
31272009 & 2008-12-20 & 4.22 & $115^{+4}_{-5}$  &  $1.2\pm 0.2$ & $1.3\pm 0.3$ & $8.8\pm 1$ & 1.9 (9 dof)\\
31272012 & 2009-01-10 & 3.71 & $118\pm5$  &  $1.3\pm 0.2$ & $1.5\pm 0.3$ & $9.5\pm 2$ & 1.1 (7 dof)\\
31272013 & 2009-01-16 & 4.16 & $116^{+4}_{-5}$  &  $1.3\pm 0.2$ & $1.4\pm 0.2$ & $9.2\pm 2$ & 1.2 (8 dof)\\
31272014 & 2009-01-23 & 1.45 & $113^{+8}_{-12}$  &  $1.1\pm 0.4$ & $1.2\pm 0.4$ & $8.1\pm 3$ & 0.1 (1 dof)\\
31272015* & 2009-01-30 & 3.95 & $116^{+4}_{-5}$  &  $1.2\pm 0.2$ & $1.4\pm 0.2$ & $9.0\pm 1$ & 0.8 (8 dof)\\
\hline
\end{tabular}
\label{tab:swift}
\begin{tablenotes}
\item[]Note. -- The quoted errors represent 90\% confidence levels. 
$F_{\mathrm{X}}$ represents the 0.5-10~keV total model flux (described in the text) and $F_{\mathrm{X}}^{\mathrm{th}}$ gives the 0.01-100 keV \texttt{NSATMOS} flux (both unabsorbed and in units of $10^{-12}~\mathrm{erg~cm}^{-2}~\mathrm{s}^{-1}$). $L_{\mathrm{X}}$ gives the 0.01-100 keV luminosity of the \texttt{NSATMOS} model component (assuming a source distance of 7.4~kpc). The exposure times of observations marked with an asterisk were corrected for (parts of) eclipses.
\end{tablenotes}
\end{threeparttable}
\end{table*}

\subsection{Swift data} 
In addition to the \chan data, we obtained \textit{Swift}/XRT TOO observations of \exo returning to quiescence (see Table~\ref{tab:swift} for an overview). The XRT data, collected in the Photon Counting mode, were processed using standard \swift analysis threads.
We extracted source spectra (using \textit{Xselect} v. 2.3) from a circular region with a radius of 15\arcsec, while background spectra were obtained from an annular region with an inner (outer) radius of 50\arcsec\ (100\arcsec). The spectra were grouped to contain bins with a minimum number of 10 photons. We reduced the exposure times of those observations that contained eclipses according to the ephemeris of \citet[][]{wolff2008c} to calculate the correct non-eclipse time-averaged fluxes (see Table~\ref{tab:swift}). 

We fitted all grouped \swift spectra with a combined \texttt{NSATMOS} and powerlaw model, were we fixed all parameters except the effective temperature. Different fits to the \chan spectral data yield similar $\chi^2$ values, so there is no preferential model to use for the \swift data (see Table~\ref{tab:spec}). We picked the fit with $D=7.4$~kpc (the best distance estimate), $\Gamma=1$, $M_{\mathrm{NS}}=1.4~\mathrm{M_{\odot}}$, $R_{\mathrm{NS}}=17.1~\mathrm{km}$ and $N_{\mathrm{H}}=0.12\times 10^{22}~\mathrm{cm}^{-2}$. The powerlaw normalization was adjusted for each observation so that this component contributes $17\%$ of the total 0.5-10 keV flux. To improve the statistics, we tied the spectral parameters between observations 31272003 and 31272004, which were performed only one day apart. The results are presented in Table~\ref{tab:swift}. 

Figure~\ref{fig:temp} displays the effective temperatures and thermal bolometric fluxes derived from the \chan and \swift data. The \chan observations (obtained  $<3$ days apart) are plotted as a single data point (with an error on the time to indicate the spread of the two observations). The \swift observations 31272003 and 31272004 are also plotted as a single point.
The bottom panel displays the evolution of the effective temperature of \exo together with the data points and curve fits of KS~1731--260 and MXB~1659--29 \citep[taken from][]{cackett2006,cackett2008}. \citet{cackett2006} set the reference time, $t_{0}$, for KS~1731--260 and MXB~1659--29 to the day of the last detection with \textit{RXTE}/PCA. For \exo we set $t_{0}$ at 2008 September 5, which is in between the last detection with \textit{RXTE}/PCA (August 12) and the first \swift observation (September 28).

\section{Discussion}
We obtained two \chan and twelve \swift observations within five months after the cessation of the very long ($\sim$24 yrs) active period of EXO~0748--676. 
We found (assuming a neutron star atmosphere model \texttt{NSATMOS}) a relatively hot and luminous quiescent system with a temperature of $kT^{\infty}_{\mathrm{eff}} \sim 0.11-0.13$~keV and a thermal bolometric luminosity of $\sim (8.1-16) \times 10^{33}~(\mathrm{d/7.4~kpc})^2~\mathrm{erg~s}^{-1}$. 
In addition to a soft, thermal component, the \chan data reveal a hard powerlaw tail, which contributes $\sim 20\%$ to the total 0.5-10 keV luminosity of $8.5 \times 10^{33}~(\mathrm{d/7.4~kpc})^2~\mathrm{erg~s}^{-1}$. 

\begin{figure}
 \begin{center}
          \includegraphics[width=7.8cm]{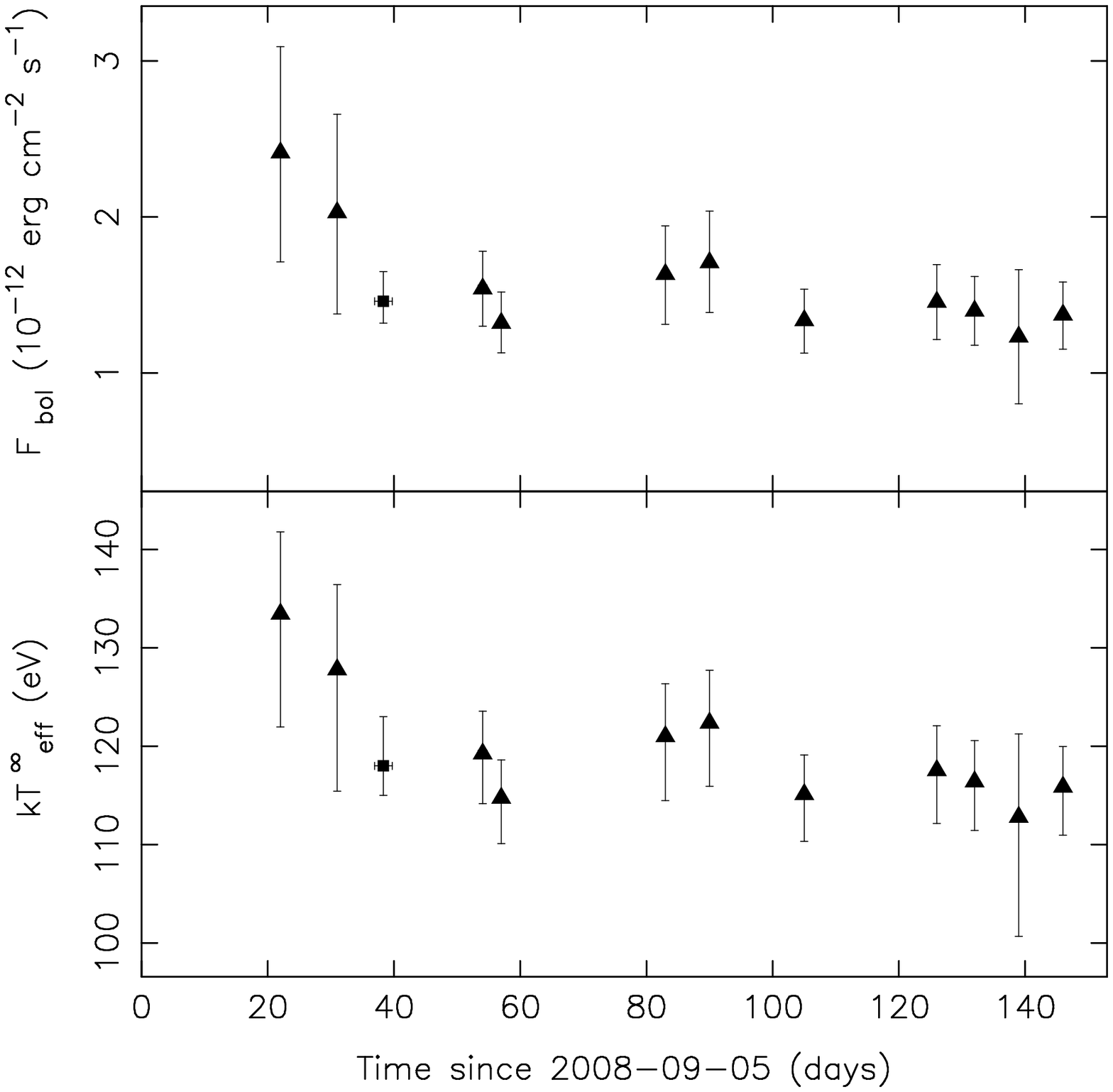}\vspace{0.1cm}
    \includegraphics[width=7.8cm]{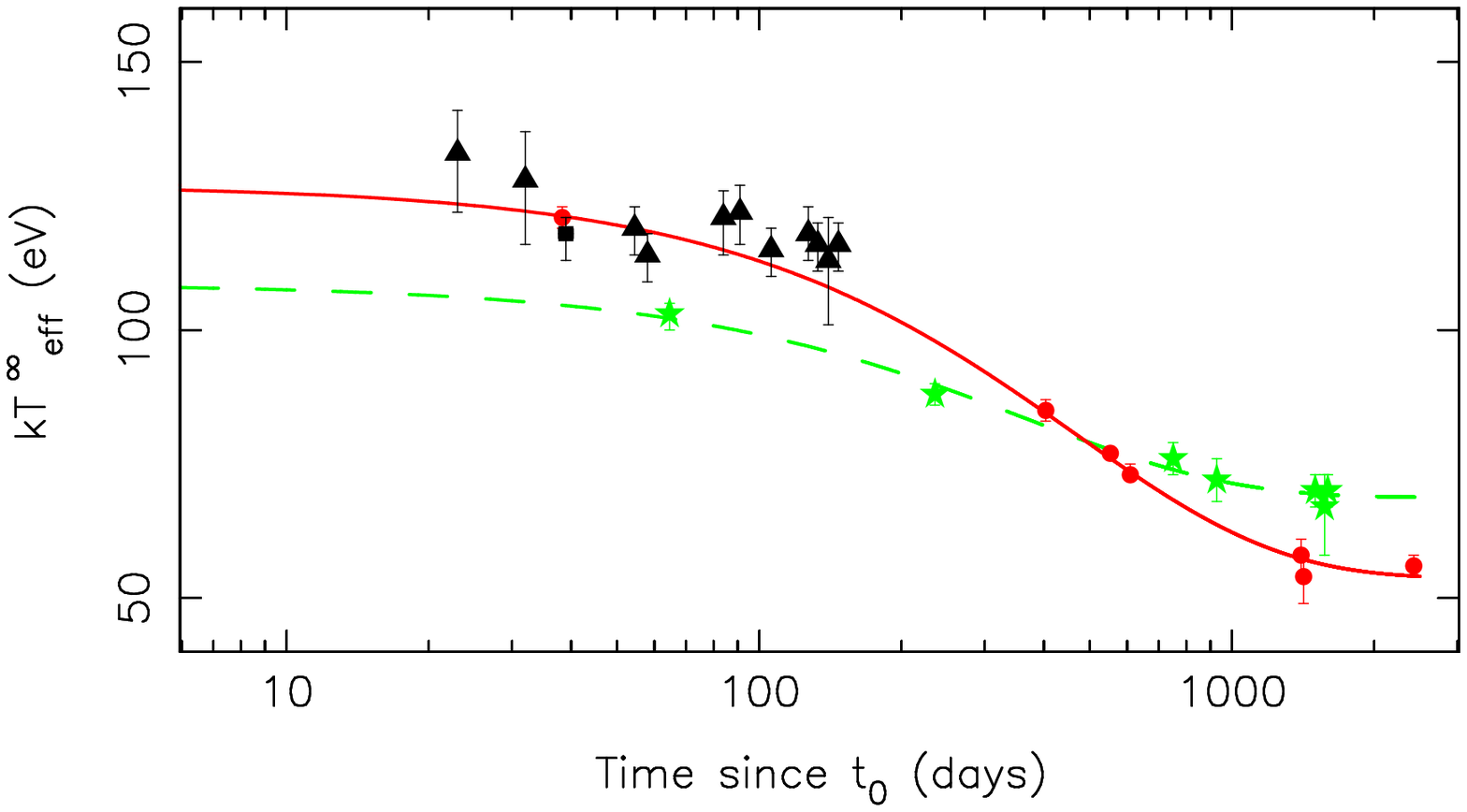}
    \end{center}
\caption[]{Evolution of the bolometric flux (top) and effective temperature (middle/bottom) of EXO~0748--676, deduced from \textit{Chandra}/ACIS-S (squares) and \textit{Swift}/XRT (triangles) observations. The bottom panel displays the effective temperatures of KS~1731--260 \citep[green stars; from][]{cackett2006} and MXB~1659--29 \citep[red bullets; from][]{cackett2006,cackett2008}, in addition to the data points of EXO~0748--676. The exponential decay fits to the data of KS~1731--260 and MXB~1659--29 are also shown (dashed green line and solid red line respectively).
}
 \label{fig:temp}
\end{figure}

Comparing the evolution of the effective temperature of \exo with that of KS~1731--260 and MXB~1659--29 (bottom panel Figure~\ref{fig:temp}) illustrates that the current data of \exo is consistent with the fit through the data of MXB~1659--29 (as well as with KS~1731--260 if the temperatures would be scaled). This suggests that the neutron star crust may thermally relax in the coming years, revealing a cooling curve as has been observed for KS~1731--260 and MXB~1659--29. The current data set can then provide an unique insight into the early stages of neutron star cooling, and can possibly put constraints on the amount of heating in the outer crustal layers \citep[][]{brown08}. 

However, the top and middle panel of Figure~\ref{fig:temp} suggest that the effective temperature and thermal bolometric flux of \exo have not decreased during the past three months (see also Table~\ref{tab:swift}). The 0.5-10 keV luminosity remains approximately constant at $L_{\mathrm{X}}\sim 8 \times 10^{33}~(\mathrm{d/7.4~kpc})^2~\mathrm{erg~s}^{-1}$, which is close to the value deduced from an \textit{Einstein} observation in 1980 (see Section~\ref{subsec:exo}).
There are several explanations that can account for the current high luminosity of EXO~0748--676 and are consistent with the \textit{Einstein} detection of \exo in 1980.

Firstly, we cannot exclude the possibility that we detect low-level accretion from EXO~0748--676, since the resulting radiation spectrum may have a shape similar to that expected from crustal heating \citep[e.g.,][]{zampieri1995}. We made Fast Fourier Transforms of the \chan data (excluding the eclipses), but did not find any short-timescale ($< 10^{4}$~sec) variability that might indicate continued accretion \citep[][]{rutledge2002_aqlX1}. 

If the observed thermal emission is due to crustal heating, then the constant luminosity might imply that the crust and core have already reached thermal equillibrium. 
The neutron star in \exo would then be relatively hot compared to other quiescent systems \citep[see e.g., figure 4 of][]{heinke2008}. Such a high quiescent luminosity can be explained by standard core cooling. Since enhanced neutrino emission mechanisms are suppressed only when the density in the core is relatively low, this scenario would imply that the neutron star in \exo is not very massive and has not had enough time to accrete a significant amount of matter (the exact mass limit for enabling enhanced core cooling mechanisms is model dependent). 
 
A high time-averaged mass-accretion rate can also give rise to a high quiescent luminosity. \citet{parmar1986} stated that between 1970 and 1980 no outburst reaching $\sim 10^{36}~\mathrm{erg~s}^{-1}$ was observed for \exo using \textit{Uhuru, Ariel V} and \textit{HEAO-1}, indicating that in the 10 yrs prior to the \textit{Einstein} detection the source was in quiescence (at least, no similar long outburst as the most recent one occurred; shorter outbursts of weeks or even months cannot be excluded). Besides this, we cannot put any additional constraints on the duty cycle of EXO~0748--676. Normally, X-ray transients reside significantly longer in quiescence than in outburst, but for \exo the quiescence state might be similar in duration to the outburst episodes. The neutron star core temperature could then be maintained by repeated accretion episodes at a significantly higher level than would be the case if it would spend most of its time in quiescence. 

Furthermore, a high quiescent luminosity can be accounted for if the neutron star crust has a low thermal conductivity, so that it will cool on a time scale of decades rather than a few years and remains hot for a long time \citep[][]{rutledge2002,shternin07}. A drawback of this explanation is that there is no obvious reason why the neutron star in \exo would be so different in this respect from KS~1731--260 and MXB~1659--29, for which a low crust conductivity can be ruled out \citep[][]{shternin07,brown08}. This would also oppose independent molecular dynamics simulations that predict a regular crystal lattice structure \citep[][]{horowitz2007}. 
 
More \chan observations of \exo are scheduled for this year and these will provide insight into the different scenarios discussed above. 

\vspace{-0.2cm}

\section*{Acknowledgments} 
We are grateful to the referee, Nathalie Webb, for very useful comments. This work was supported by NWO. 
We acknowledge the use of the \textit{Swift} public data archive. EMC was supported by NASA through the Chandra Fellowship Program. MTW, PSR and KSW acknowledge the United States Office of Naval Research. J.H.\ and W.H.G.L.\ gratefully acknowledge support from Chandra grant GO8-9045X.

\bibliographystyle{mn2e}

\label{lastpage}
\end{document}